# PARTICLE-MESH METHODS ON THE CONNECTION MACHINE

ROBERT FERRELL

*Thinking Machines Corporation, 254 First Street, Cambridge, MA 02142,*

and

EDMUND BERTSCHINGER

*Department of Physics, MIT Room 6-207, Cambridge, MA 02139*

We describe an efficient Particle-Mesh algorithm for the Connection Machine CM-5. Our particular method parallelizes well and the computation time per time step decreases as the particles become more clustered. We achieve floating-point computation rates of 4–5 MFlops/sec/processing node and total operations (the sum of floating-point and integer arithmetic plus communications) of 5–10 MOps/sec/processing node. The rates scale almost linearly from 32 to 256 processors. Although some of what we discuss is specific to the CM-5, many aspects (e.g., the computation of the force on a mesh) are generic to all implementations, and other aspects (e.g., the algorithm for assignment of the density to the mesh) are useful on any parallel computer.

## 1. Introduction

Particle-mesh methods are used to compute long-range forces in a system of self-interacting particles by solving the field equations on a mesh or grid. Also known as particle-in-cell methods, these algorithms find widespread application in plasma physics and astrophysics.[1,2] The advantage of these methods is that pair forces on $N$ particles are computed in $O(N)$ or $O(N \log N)$ operations rather than $O(N^2)$. However, the spatial resolution of the force field is limited by the grid. When higher resolution is required in molecular dynamics computations the pair force may be split into short- and long- (or medium-) range parts,[2,3] with Particle-Mesh (hereafter, PM) used for the latter.

The aim of the PM method is to compute particle accelerations by solving a linear field equation relating the acceleration $\vec{g}$ and density (of, e.g., mass or charge) $\rho$. We illustrate with the Poisson equation for gravity (with Newton's constant $G = 1$):

$$\vec{\nabla} \cdot \vec{g} = -4\pi\rho \ , \quad \vec{g}(\vec{x}) = \int d^3x' \, \frac{\rho(\vec{x}\,')(\vec{x}\,' - \vec{x})}{|\vec{x}\,' - \vec{x}|^3} \ . \tag{1}$$

The method is not restricted to the Coulomb interaction but works for any problem where the force field is a sum over particles or, equivalently, a linear convolution of the density. The convolution may be performed rapidly in the Fourier domain using the Fast Fourier Transform (FFT) algorithm. Other rapid algorithms exist for evaluating pair Coulomb forces,[2,4] but Fourier convolution has the advantage of working for any linear field equation and in practice it is suitably fast.

The force calculation in the PM method may be divided into three phases:

1. Compute the density on a grid by interpolating from particle positions.

2. Compute the potential or force on the grid from the density using Fourier transform (or other) techniques.

3. Interpolate the force back to the particles.

There are many ways to accomplish each phase, several of which are described in Refs. 2 and 3.

In this paper we will first describe a new algorithm for Phases 1 and 3 which is efficient on parallel computers such as the CM-5 which we are using. We will then discuss Phase 2, with particular emphasis on the FFT solution of the Poisson equation on the CM-5. We use an anti-aliasing filter to minimize grid artifacts (this procedure is called "Quiet PM" in Ref. 2). In Appendix A we give a detailed account of how we construct the optimal filter.

Our notation is similar to that of Ref. 3. The number of particles is $N$ and the number of grid cells along one dimension is $M$. We assume a cubical grid in three dimensions although it is easy to generalize to a rectangular grid in any number of dimensions. The unit of length is taken to be the grid spacing. The mean mass per grid cell is defined to be unity, so the total mass in the cube is $M^3$. The vertices of the grid have positions given by the integer triples $\vec{n} = (n_1, n_2, n_3)$, with $0 \leq n_1, n_2, n_3 < M$.

## 2. Programming the Connection Machine

The Connection Machine CM-5 is a distributed memory, parallel processing computer built with tens to thousands of processing nodes. Each node of the CM-5 has 32 MBytes of memory, a Sparc microprocessor and 4 vector processor accelerators. The nodes are connected by a data network and a control network. The data network is used to send pieces of data from any node to any other node, as required for gather or scatter operations, for instance. The control network is used to send data from a single node to all other nodes, as required, for instance, for broadcasting a number from one node to all other nodes. The control network is also used to synchronize the nodes.

Efficient use of the CM-5 (or any distributed memory computer) demands that the program exploit data locality as much as possible. This means that the algorithms used must be such that each processing node references data on that node most of the time, and only moves data between nodes occasionally. Furthermore, the best performance is obtained when most of the nodes have about the same amount of work to do. If that is the case, the algorithm is load balanced. The PM algorithm we describe below has both of these desirable properties, and therefore makes efficient use of the CM-5's computing power.

A useful paradigm for programming a parallel computer such as the CM-5 is the Data Parallel programming model. In the Data Parallel model, one imagines that each data element (array element) has an associated processor which does the computational work on that element. Since in general there are many more data elements than processors, in practice we associate each data element with a "virtual processor." The compiler and system software map the virtual processors onto the physical processors. For a systematic description the reader is referred to the paper by Hillis and Steele.[5]

The Data Parallel model provides a framework for development of efficient algorithms. In many physics simulations, the laws of physics are specified in local terms. For a computer simulation, this means that algorithms written in the Data Parallel model automatically have each processing node computing mostly on data which are stored on that node. Furthermore, since the laws of physics are the same everywhere, each data element, or virtual processor, is doing the same amount of work. This means many Data Parallel algorithms are both local and load balanced by construction. This is the power of the Data Parallel paradigm. We will use the Data Parallel paradigm for our PM algorithm.

### 3. Issues for an Efficient Algorithm

In a PM simulation there are two fundamental data structures. The first is a list of particle positions (and other information needed about the particles, such as their velocities). This list is usually stored as a one dimensional array (or $d$ one-dimensional arrays in $d$ dimensions). The second data structure is a mesh. The mesh has the dimensionality of the simulation space ($d = 3$ in our case), and is typically stored as an array of that many dimensions. There may be different numbers of mesh cells than particles.

Following the dictates of the Data Parallel paradigm, we assign each particle to a virtual processor. These virtual processors are then mapped to the physical processors. On the CM-5 this is a linear mapping since the particle list is one dimensional. If there are $N$ particles and *NProc* processors, then each physical processor simulates $vpr = \lceil N/NProc \rceil$ virtual processors, where *vpr* is called the *virtual processor ratio*. The relation between particle $n$ and processor $P$ is $P = \lfloor (n-1)/vpr \rfloor + 1$. For higher dimensional arrays the distribution of virtual processors onto physical processors is still governed by the virtual processor ratio, but the relation between virtual processor number and physical processor number is more complicated. Besides mapping the particles, we will also assign each mesh cell to a virtual processor, and then these are mapped to the physical processors.

In general there is no correlation between the processor storing a particle's position and the processor storing the mesh cell which contains that particle. This means that our algorithm will have a non-local component because we will have to move particle data between processors in both the density assignment and the force interpolation phases.

Furthermore, depending on the distribution of the particles, it may be that some mesh cells have many more particles in them than do others. This could present a load balancing problem. The most important feature of the algorithm we will describe later is that it is load balanced for all density distributions.

### 4. Assignment of Density to Mesh: Naive Parallelism

For simplicity, we consider a Nearest Grid Point (NGP) scheme.[2] The technique we describe is easily extended to higher order interpolation schemes (Section 9). In the NGP scheme, the (discrete) density is the array whose value at each grid point is the sum of all the masses of the particles nearest to that grid point.

A simple numerical scheme for implementing this is:
```
For each particle:
    Compute the NGP
    Add mass of particle to NGP
```
Clearly the first step can be done for all the particles in parallel. If no two particles share an NGP, then it is clearly possible to parallelize the second step over the particles. Parallelism is possible even in the case where more than one particle is contributing to the mass at a given grid point. Readers familiar with vector processors will realize that this step does not vectorize.

Using the Data Parallel programming model, we instruct the virtual processor associated with each grid point to sum all mass contributions to that grid point. This is a reasonable solution only because in our paradigm each data element has associated with it a virtual processor.

One way to implement this phase (either explicitly by the user, or else implicitly by the system software) is to first send all particle masses to queues at their respective destination grid points. Then, after all the masses have been delivered to the queues, the virtual processor at each grid point executes a sum over all entries in its queue.

If the density is nearly uniform, this is an efficient technique. In that case, all grid cell queues receive nearly the same number of masses to sum together. Furthermore, the wires which carry data from one processor to another are nearly uniformly loaded.

However, it is also clear that if the density is concentrated in just a few clusters, then the load on the machine will not be uniform. In particular, the virtual processor (and the physical processor it is assigned to) representing one of the grid points in a dense cluster will have to do a lot of work, while processors representing empty regions of space will have no work. In addition, the wires leading to the heavily loaded processors will be clogged with messages, while other wires will be completely unused. This is a classic load balancing problem, which apparently can become arbitrarily bad, in the sense that the time to complete the density assignment can grow arbitrarily large.

In subsequent sections we will introduce an algorithm which is load balanced. That algorithm requires us to use some *Parallel Prefix Operations*. We introduce these operations in the next section, then in Section 6 we describe how they are used in a load balanced mesh assignment algorithm, and in Section 7 we describe how they are used in a load balanced force interpolation algorithm.[a]

## 5. Parallel Prefix Operations

Parallel prefix operations, also referred to as *Scans*, are a method of turning certain kinds of global communications into regular, mostly local, communications. Figure 1 shows a Scan with Add, used to compute a running sum of a list of numbers. Notice that the number of inter-processor communication operations is $O(\log_2 NProc)$. This is much better than the $O(NProc)$ which one might naively have assumed were required. Scans can be formed for any binary associative operator (the operator does not have to be commutative). Scan operations can be either upward or downward. Figure 1 shows a Scan Add Downward. Mathematically, a Scan Add Downward on the vector $A$ is expressed as: $B_i = \sum_{j=N}^{j=i} A_j$, $i \in [1, N]$. A Scan Add Upward is: $B_i = \sum_{j=1}^{j=i} A_j$, $i \in [1, N]$.

---

[a] It is possible to vectorize these parallel prefix operations.[6] Consequently, the algorithm we describe in Section 6 can be used to vectorize the density assignment step discussed in Section 4.

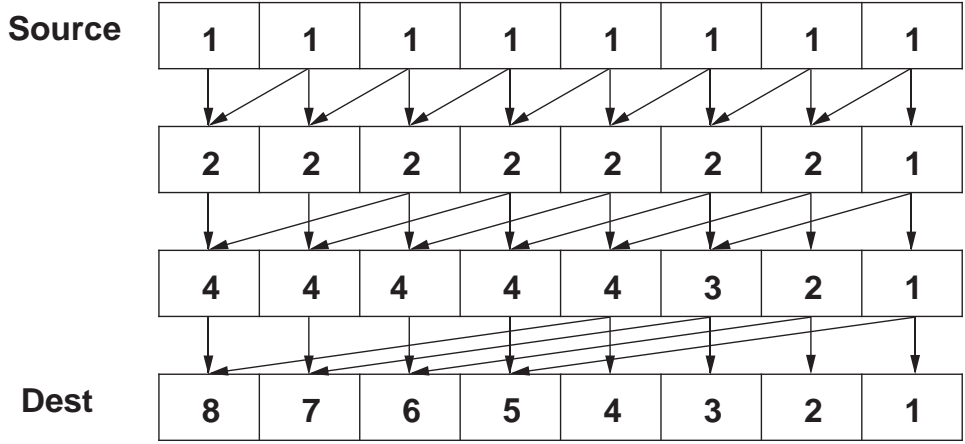

Figure 1: *The communication pattern for a Scan Add on 8 processors. Coincidence of two arrays means add the two numbers from which the arrows originate. For NProc processors, the Scan operation takes $\log_2$ NProc communication steps.*

In our algorithm we require a more general version of parallel prefix operations called *Segmented Scans*. Figure 2 shows a Segmented Scan Add Downward. An auxiliary list of logical values is used to divide a linear array (list) into segments. A .TRUE. indicates the start of a new segment. The Segmented Scan Add computes a running total within each segment. The Segmented Scan Add can also be completed with $O(\log_2 NProc)$ communication operations. The reader is referred to Ref. 5 for a complete description of the Segmented Scan algorithm.

Segmented operations can also be performed for any binary associative operator. In our PM algorithm, in addition to a Segmented Scan Add we will need a Segmented Scan Copy. Figure 3 shows the results of a Segmented Scan Copy Upward. The first element of every segment is copied to all other elements in that segment.[b]

## 6. Assignment of Density to Mesh for Clustered Distributions

We now discuss a solution to the load balancing problem introduced at the end of Section 4. In this section we present a new algorithm for density assignment which actually speeds up as the particles become more clustered. For clustered particle distributions, or even uniform ones with multiple particles at each grid point, the method described in this section is faster than the naive algorithm presented in Section 4.

---

[b] A Segmented Scan Copy can be written in terms of a Segmented Scan Add, by preceding the scan add with `WHERE(.NOT. Segment) Source = 0.0`.

| Source | 1 | 1 | 1 | 2 | 3 | 3 | 3 | 3 |
| Segment | T | F | F | T | T | F | F | F |
| Dest | 3 | 2 | 1 | 2 | 12 | 9 | 6 | 3 |

Figure 2: *An example of a Segmented Scan Add Downward. This is similar to a Scan Add, except that a new running total is started at the beginning of each segment. This operation is a bit more costly than a simple Scan Add, but still completes in $O(\log_2 NProc)$ communication steps. This operator is required for the assignment of density to the mesh.*

| Source | 1 | 2 | 3 | 4 | 5 | 6 | 7 | 8 |
| Segment | T | F | F | T | T | F | F | F |
| Dest | 1 | 1 | 1 | 4 | 5 | 5 | 5 | 5 |

Figure 3: *An example of a Segmented Scan Copy Upward. The first value in each segment is copied to all other elements in each segment. This operator is required for the interpolation of the force back to the particles.*

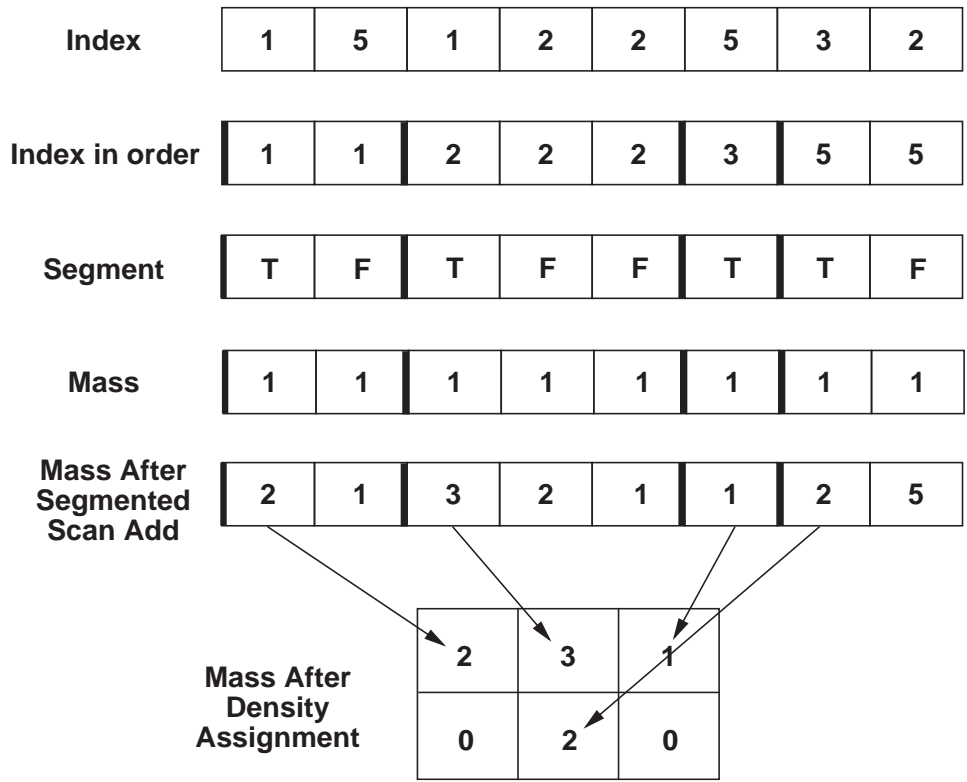

Figure 4: *The algorithm for assigning the density to the mesh. The key to the algorithm is the Segmented Scan Add for totaling all the masses in each cell. This step, which a priori seems sequential, can be done in parallel in logarithmic time. The final send of the mass to each mesh cell is load balanced because each mesh cell receives at most one message.*

A load imbalance arises when some grid points, and consequently some physical processors, have to do significantly more additions than the other grid points. The way to develop a load balanced algorithm is to assure that all the processors are doing an equal number of additions. The cost of this method is that the particle list must be ordered in some way.

Figure 4 shows the algorithm. The particle list is ordered so that particles in the same mesh cell are contiguous in the list. At that stage, the particle list is a set of segments, and an auxiliary logical array `Segment` is constructed to mark the segments.[c] Within each segment, all particles have the same nearest grid point.

---

[c]The construction of the list `Index` is required so that we can order the particle list. The

Within each segment we add together the masses of all the particles in that segment. We then send that number to the corresponding grid point. Evidently, each grid point receives at most one number, so there is no problem with load balancing. The addition of all the masses in a segment is done using the Segmented Scan Add. This method is fast, as noted in Section 5, and uses all the processors equally because the Scan is performed on all particles. Furthermore, since the number of messages sent has been reduced (one message per occupied mesh cell now, rather than one message per particle), the load on the network wires is reduced, so the messages are delivered more quickly. Thus the run time is reduced as the particles become more clustered.

The efficient density assignment algorithm is summarized as follows:

```
For each particle:
       Compute the NGP (mesh indices I,J,K)
       Compute Index = I+(J-1)*M+(J-1)*(K-1)*M*M
Order the particles according to Index (any ordering is fine)
Construct Segment(i) = Index(i) .NE. Index(i-1)
For each segment:
       Sum masses of all particles in that segment
              (using Segmented Scan Add)
       Send the accumulated mass to the NGP
```

### 7. Interpolation of the Force to the Particles

This phase, the last stage of the PM force computation, is the inverse of the assignment of the density to the mesh. By this stage, we have constructed a force field at each point on the mesh by methods discussed below, and we have to interpolate from that a force on each particle. Again this requires moving data between the two fundamental data structures, the mesh and the particle list.

Momentum conservation requires that the same interpolation scheme used for assigning the density to the mesh be used to assign the force back to the particles.[2] For our example we are using NGP, so the force on a particle is simply the mesh force at the nearest grid point.

A simple numerical scheme for implementing this is:

```
For each particle:
       Compute the NGP
       Get the force at that NGP
```

This algorithm is clearly parallelizable because each particle gets one and only one force from the mesh — there are no collisions at the destination (the particle list). For this reason it has long been recognized that it is possible to vectorize this step, but not the density assignment step.

However, this naive implementation still suffers from a load balancing problem. To understand this, remember that on a distributed memory computer, moving data between different data structures requires moving data between different physical processors. During the force interpolation phase, each particle gets a force value from the virtual processor representing its NGP. This is implemented in two steps. First the virtual processor representing each particle sends a message to its NGP

---

construction of Index from the grid indices I,J,K is not unique.

requesting the force value. Next, the virtual processor representing the NGP sends back the force value. If the density distribution is inhomogeneous a particular virtual processor representing a mesh cell in a high density region will have to receive and reply to many more requests than a virtual processor in a low density region. That means that some *physical* processors will have a lot more work to do than others.

A load balanced algorithm for the interpolation of the force to the particles is illustrated in Figure 5. It is very similar to the algorithm for assigning the density to the mesh. In fact, the first part is identical to what is done in Phase 1. (Since the particles do not move between Phase 1 and Phase 3 in our implementation we skip the ordering step in Phase 3.) Our algorithm is thus:

```
For each particle:
        Compute the NGP (mesh indices I,J,K)
        Compute Index = I+(J-1)*M+(J-1)*(K-1)*M*M
Order the particles according to Index (any ordering is fine)
Construct Segment(i) = Index(i) .NE. Index(i-1)
For each segment:
        Get the force from the NGP (one per segment)
        Copy the force to all other particles in that segment
                (using Segmented Scan Copy)
```

Since there is only one get per segment, each mesh cell services at most a single request. Therefore this step is well load balanced. (For very inhomogeneous distributions this algorithm dramatically reduces the number of messages that must be transmitted — one per occupied mesh cell rather than one per particle. The load on the communication network is likewise reduced, so this algorithm is significantly faster for inhomogeneous distributions than for homogeneous distributions.) In addition, since the Segmented Scan Copy uses all the processors of the computer equally, this step is also load balanced. The computations in Phase 1 are completely load balanced and the computation of the force on the grid (Phase 2) is load balanced, so the whole algorithm is load balanced. Consequently we have accomplished our goal of developing a PM algorithm which does not slow down as the particles become more clustered.

## 8. Solving the Poisson Equation

Now that we have given algorithms for efficiently computing the density from a list of particles and then interpolating a field defined on a mesh back to the particles, we must address the intermediate phase: computing the force field from the density. Both objects are defined on the mesh, so the only data motion involved is that required to solve the field equation. We assume here that the field equation is Eq. 1.

We are interested in periodic boundary conditions, for which the Fast Fourier Transform (FFT) algorithm provides an efficient way to solve Eq. 1 on a mesh. (One could still use FFTs to solve the Poisson equation with vacuum or conducting boundary conditions, at the expense of increased storage and/or extra FFT calls. See Ref. 2 for examples of this and alternative solution methods.)

We introduce the potential $\phi$ related to the force by $\vec{F} = -\vec{\nabla}\phi$. (We use $\vec{F}$ and $\vec{g}$ interchangeably because the gravitational charge/mass ratio is unity. For electric forces $\vec{F}$ would be replaced by the electric field $\vec{E}$.) Fourier transformed quantities are written with a caret and are evaluated on the reciprocal lattice $\vec{k} = (k_1, k_2, k_3)$. We choose units so that the wavevector components are integers bounded by the Nyquist frequency, $-M/2 \leq k_1, k_2, k_3 < M/2$.

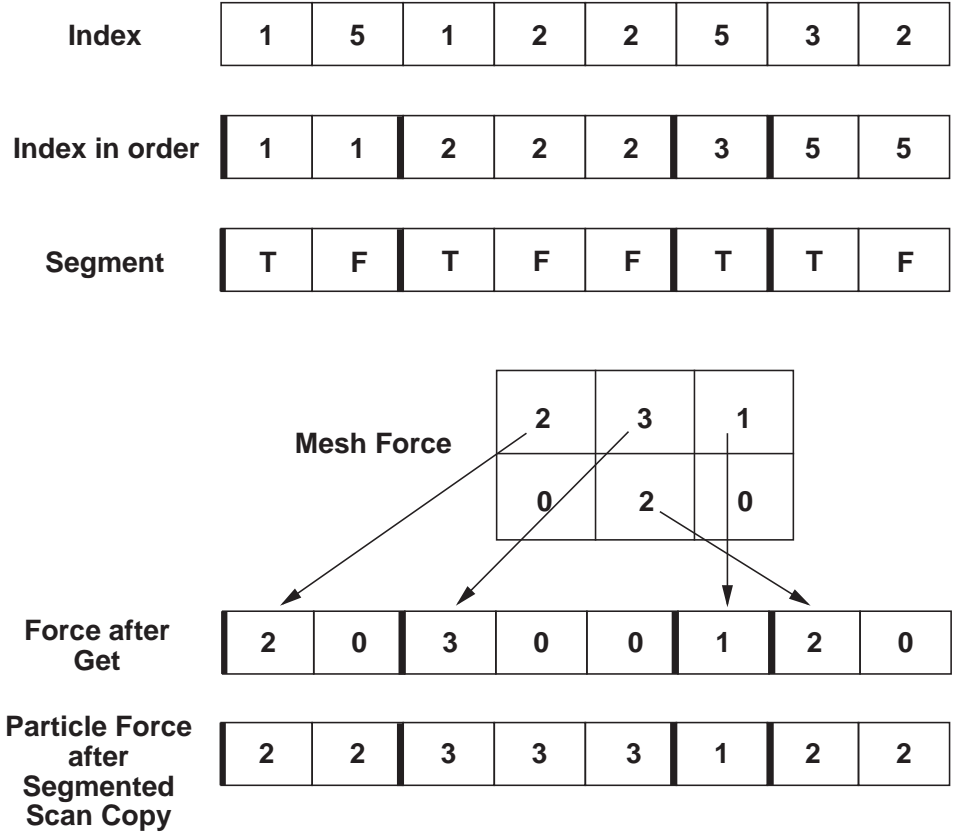

Figure 5: *The algorithm for interpolating the mesh force to the particles. This algorithm is load balanced because each mesh cell sends at most one message. The Segmented Scan Copy spreads that value to all particles in the same mesh cell. This is an efficient algorithm because the Segmented Scan Copy can be fully parallelized, as explained in the text.*

Including a filter $T(\vec{k})$, the solution to the Poisson equation for the potential in the spectral domain is

$$\widehat{\phi}(\vec{k}) = -4\pi \left(\frac{M}{2\pi}\right)^2 T(\vec{k}) \frac{\widehat{\rho}(\vec{k})}{k^2} \ . \tag{2}$$

The factor $(M/2\pi)^2$ is present to give units to $k^{-2}$; recall that our mesh has length $M$. As described in Appendix A, we apply a filter $T$ so as to minimize the aliasing errors introduced by discretizing the density onto a mesh.

After computing the potential we compute the total potential energy,

$$U = -\sum \widehat{\rho}^* \widehat{\phi} \ , \tag{3}$$

where the sum is taken over the reciprocal lattice. We then compute the force field $\vec{F} = -\vec{\nabla}\phi$ in Fourier space. The gradient may be approximated either by finite differences in the spatial domain or by a gradient operator in the spectral domain:

$$\widehat{\vec{F}}(\vec{k}) = -\left(\frac{2\pi}{M}\right) i\vec{k}\,\widehat{\phi}(\vec{k}) \ . \tag{4}$$

We prefer this spectral operator method, even though it requires more FFT calls, because its wider spatial frequency response leads to more accurate forces. However, because $\vec{F}(\vec{x})$ is real, we must set the normal component of $\widehat{\vec{F}}(\vec{k})$ to zero on the surfaces of the fundamental Brillouin zone, $k_q = \pm M/2$ for components $q = 1, 2, 3$.

In summary, our procedure for obtaining the forces is

```
Compute density on the mesh
Transform to spectral domain using FFT
Multiply by Poisson operator to get transform of potential
For each component of the force:
    Multiply potential by gradient operator
    Transform back to spatial domain using FFT
    Interpolate force to particles
```

Each point of the reciprocal lattice is mapped to a virtual processor so that the multiplication steps are Data Parallel operations. However, parallelizing the FFT algorithm requires more work.

A description of the FFT algorithm used on the CM-5 is given the Connection Machine Scientific Subroutine Library documentation.[7] In developing a parallel FFT, the most important consideration is to keep the amount of time spent moving data between processing nodes to a minimum. Since the FFT is a global algorithm (all data elements communicate with all other data elements), there is no way to eliminate inter-processor communication entirely. On the CM-5, for most problem sizes the most efficient algorithm splits the FFT into a communication phase and a computation phase.

Assume the mesh is distributed across the processors in some arbitrary fashion. Then the FFT algorithm in three dimensions is as follows.

```
Rearrange data so that for each J,K,
    all of F(I,J,K) are on a single processor for all I
FFT the first dimension of F
Rearrange data so that for each I,K,
    all of F(I,J,K) are on a single processor for all J
FFT the second dimension of F
Rearrange data so that for each I,J,
    all of F(I,J,K) are on a single processor for all K
FFT the third dimension of F
Rearrange data to original order
```

This method effectively separates the computation from the communication. Since the communication is the most expensive part, it is desirable to limit this as much as possible. For this reason we organize our data so that, as a matter of course, the first dimension of the mesh array $F$ to be transformed is contained in a single processor. This is done using CMF$LAYOUT directives for the CMFortran compiler. This helps in three ways. First of all, the first rearrangement of the data is eliminated. Second, there is a fast algorithm for swapping an entirely on-processor dimension with a parallel dimension. That means the final communication step (and all intermediate steps, of course) can use this fast algorithm.

The third benefit we gain from making the first dimension entirely on-processor comes about because in real space $F$ is real. Therefore, we can pack it into a complex array of half the size, $F(M, M, M) \rightarrow FC(M/2, M, M)$ where $F$ is REAL and $FC$ is COMPLEX, in the FORTRAN sense of those words. This packing enables a saving of a factor of two in storage and computation.[8] On a shared memory computer, one can simply assume $F$ and $FC$ are different names for the same memory locations. On a distributed memory computer, that is not necessarily the case. However, since the first dimension of $F$ (and $FC$) is on-processor, the packing step

```
DO I = 1,M/2
    FC(I,:,:) = CMPLX(F(2*I-1,:,:),F(2*I,:,:))
ENDDO
```

does not require any interprocessor communication, and consequently takes very little time compared to the rest of the algorithm. This would not be true if the first dimension were distributed among multiple processors. After transforming from the spectral to the spatial domain the data must be unpacked by the inverse procedure:

```
DO I = 1,M/2
    F(2*I-1,:,:) = REAL(FC(I,:,:))
    F(2*I,:,:) = AIMAG(FC(I,:,:))
ENDDO
```

## 9. Higher-Order Interpolation Schemes

The use of a mesh to solve the Poisson equation introduces errors relative to the exact solution for point-like or even smoothed particles. Force accuracy is limited by the mass assignment scheme. The zeroth-order NGP (Nearest Grid Point) scheme is simple but not very accurate. We therefore replace it with the second-order

Triangular Shaped Cloud (TSC) scheme described in Ref. 2. We describe this method here as it is used in the density computation phase of our PM algorithm. A similar procedure is used in the force interpolation phase.

In the NGP scheme, the mass of a particle is assigned entirely to its NGP. In the TSC scheme, the particle's mass is spread over a cube (in three dimensions) of 27 grid points centered on the NGP. The weight given to each of these points is the product of three weights, one for each dimension. For a particle whose first coordinate is $x$ (in units of the mesh spacing), the corresponding NGP index is $I = \lfloor x + \frac{1}{2} \rfloor$. The weights assigned to $I$ and $I \pm 1$ are

$$W_I(x) = \frac{3}{4} - (x - I)^2 , \quad W_{I\pm 1}(x) = \frac{1}{2}\left(x - I \pm \frac{1}{2}\right)^2 . \tag{5}$$

The TSC assignment can be parallelized by defining the weights as one-dimensional arrays stored with the particle positions. Instead of a single Segmented Scan Add operation, we repeat this process 27 times, once for the NGP and each of its neighbors. Referring to our procedure shown at the end of Section 6, we simply place a serial loop of length 27 around the Scan Add and Send operations. For each loop iteration we compute the offset from the NGP ($\pm 1$ or 0 for each of the dimensions), imposing periodic boundary conditions, and form the corresponding weight factor, which is then multiplied by the particle mass. We only have to construct the Segment array once, however, from which we can obtain the correct grid point by applying the appropriate offset for each dimension. The operations in each of the 27 loop iterations are fully parallel so that the total time required for the density assignment is approximately 27 times as much as for the NGP scheme excluding the computation of Index and Segment , which is done only once in both schemes. As with the NGP scheme, our TSC assignment algorithm is fully load balanced and it speeds up for clustered particle distributions.

## 10. Timing Results and Discussion

We have tested the parallel PM algorithm on a CM-5 at the National Center for Supercomputing Applications. These results are based upon a test version of the CM-5 system software (CMOST 7.2 and CMFortran 2.1-Beta.2) and are not necessarily representative of the performance of the full version of this software. The results of the CM-5 code were verified by comparing with a serial version of the code run elsewhere.

Aside from the FFT calls, the total operations count is dominated by the TSC interpolation in the density assignment and force interpolation phases. The operation counts are $303N$ Flops for the density assignment and $990N$ for the force interpolation. About three times as many integer operations are required in addition for index computation. The operation count for the FFT calls is $10M^3 \log_2 M^3$. ($N$ is the number of particles and $M^3$ is the number of mesh cells.) A detailed breakdown of the operations count is provided in Appendix B.

We ran tests with both homogeneously distributed particles and a tightly clustered particle distribution. Both of these test problems had $N = 128^3$ and $M^3 = 256^3$. The times for the runs, averaged over at least five timesteps, are reported in Table 1. (Only the time required to compute forces is included; time integra-tion and other overhead adds a small amount to the total run time.) The times were measured using the CM-5 CM_Timer routines. For each test we list the homogeneity parameter $FracOC = Noc/N$, where $Noc$ is the number of mesh cells with at least one particle in them (i.e., the number of non-vacant NGPs). We also list the total memory usage. Finally, we summarize the effective performance in terms of floating-point and integer arithmetic as well as inter-processor communication. These rates are computed by dividing the total number of floating point operations, integer operations, or data words sent, respectively, by the total run time.

|  | Clustered | | | Homogeneous | | |
|---|---|---|---|---|---|---|
|  | 64PN | 128PN | 256PN | 64PN | 128PN | 256PN |
| Density Assignment | 2.8 sec | 1.6 sec | 1.0 sec | 5.6 sec | 2.8 sec | 1.5 sec |
| FFT | 4.6 sec | 2.5 sec | 1.3 sec | 4.6 sec | 2.5 sec | 1.3 sec |
| Force Interpolation | 7.9 sec | 4.4 sec | 2.4 sec | 11.8 sec | 6.2 sec | 3.4 sec |
| Total PM | 15.8 sec | 8.7 sec | 4.9 sec | 22.4 sec | 11.7 sec | 6.3 sec |
| *FracOC* | 0.11 | 0.11 | 0.11 | 1.0 | 1.0 | 1.0 |
| Maximum Memory | 762 MB | 810 MB | 940 MB | 722 MB | 770 MB | 900 MB |
| Net MFlops/sec/PN | 5.3 | 4.8 | 4.3 | 3.7 | 3.6 | 3.3 |
| Net MIops/sec/PN | 1.1 | 1.0 | 0.9 | 5.6 | 5.4 | 5.0 |
| Net MWords/sec/PN | 0.3 | 0.3 | 0.2 | 0.5 | 0.5 | 0.4 |

Table 1: Timing statistics for two different test problems on three different size CM-5 partitions. The run time for uniformly distributed particles is longer than for clustered particle sets, as predicted. The times decrease nearly linearly with increasing numbers of processing nodes (PNs). For both test problems, $N = 128^3$ and $M^3 = 256^3$.

The net MFlops rate for the strongly clustered particle distribution is about 5 MFlops/sec/PN. For homogeneously distributed particles the rate is about 3.5 MFlops/sec/PN. The rate is higher for clustered particle distributions because less time is spent in communication. These results confirm our expectation that our PM algorithm should speed up with clustering.

The performance rates are nearly independent of the number of processors, consequently the run time decreases nearly linearly with increasing number of processors. The reason for the slightly higher rates with fewer processors is that the virtual processor ratio is larger with fewer processors. As a result less inter-processor communication is required. Most of the run time is spent in communications.

In addition to these tests we ran a large test problem with 8 times as many particles and grid points on a partition with 256 processing nodes. The problem was slightly less clustered than the clustered distribution above, having $FracOC = 0.12$. The total memory requirement was 5.32 GB. For this run the rates in Mops/sec/PN were measured to be $(4.8, 0.9, 0.2)$ for floating-point, integer, and inter-processor communication, respectively, and the total PM timestep was 38.0 sec. This performance scales as expected from running the (8 times) smaller problem on 32 (8 times fewer) processing nodes.

We expect some performance improvements in the future. For instance, work in progress should result in higher performance for the Scans. In addition, our communication rates may be increased by hand coding some of the routines. Our optimism is supported by the efficient performance of the CMSSL FFT routines. Even though these require large amounts of data motion, as we discussed in Section 8, the net MFlops rate for the FFT calls is consistently about 12 MFlops per processing node.

In conclusion, we have demonstrated a parallel scalable algorithm for the Particle-Mesh force computation. This algorithm is useful for computing pair forces in collisionless systems and plasmas where short-range force resolution is not needed. However, for many applications, including gravity, the pair-potential has a short-range component that cannot be resolved easily by a mesh. We are currently working to implement a parallel short-range algorithm similar to the parallel Verlet neighbor list method of Giles and Tamayo.[10] The long- and short-range computations may be combined in one hybrid code that should provide an efficient parallel scalable approach to the gravitational $N$-body and similar problems.

## Acknowledgments

This work was supported in part by NSF grants AST90-01762 and ASC93-181815. We thank the director of NCSA for a discretionary allocation of supercomputer time for code development and testing.

## Appendix A  Analysis of the PM Force Computation

This Appendix presents a mathematical analysis of the Particle-Mesh force computation and derives the optimal anti-aliasing filter. This material is based on Ref. 2, to which the reader is referred for more details.

**Appendix A.1.** *Exact forces*

Before describing practical PM implementations, we first summarize the solution for the potential and forces on particles using Fourier transforms. The results presented in this subsection are exact for a periodic mass distribution and they correspond to the limit of an infinitely fine PM mesh. For convenience we choose units so that the entire cube has length $M$ and mass $M^3$, but the particle positions are not discretized in any way. These exact results will provide a standard for comparison with the approximate forces resulting when a finite grid is used.

The exact density distribution for a set of $N$ discrete points is

$$\rho(\vec{x}) = \sum_{i=1}^{N} m_i \, \delta_D^3(\vec{x} - \vec{x}_i) \; , \tag{A.1}$$

where the Dirac delta function $\delta_D$ picks out each particle with position $\vec{x}_i$ and mass $m_i$. (We will take the masses all to be equal, $m_i = M^3/N$.) In practice we may work with a smoothed density field,

$$\rho_s(\vec{x}) = \int d^3x' \, W(\vec{x} - \vec{x}') \, \rho(\vec{x}') = \sum_{i=1}^{N} m_i W(\vec{x} - \vec{x}_i) \; . \tag{A.2}$$

The convolution integral is equivalent to replacing each mass point by a cloud with density profile $W(\vec{x})$, as we will do in the PM algorithm. The size and shape of the cloud are left arbitrary for the moment, except that we assume the cloud has even parity, $W(-\vec{x}) = W(\vec{x})$. Because the volume is periodic, the volume integrals in Eq. A.2 are taken over only the fundamental cube, $0 \leq x, y, z < M$, although $W(\vec{x})$ itself is periodic, so that a particle close to an edge of the cube may spill over and contribute to the density on the opposite side of the cube.

We define the Fourier transform pair by

$$\widehat{\rho}(\vec{k}) \equiv \int d^3x \, \exp\left(-i2\pi \vec{k} \cdot \vec{x}/M\right) \rho(\vec{x}) \; ,$$
$$\rho(\vec{x}) = M^{-3} \sum_{\vec{k}} \exp\left(i2\pi \vec{k} \cdot \vec{x}/M\right) \widehat{\rho}(\vec{k}) \; . \tag{A.3}$$

The volume integral is taken over the cube while the sum over wavenumbers is taken over all integer values for the components of the wavevector $\vec{k} = (k_1, k_2, k_3)$. The spatial frequencies are discrete (with the units absorbed by the fundamental spatial frequency $2\pi/M$) because $\rho(\vec{x})$ is periodic. Because $\rho(\vec{x})$ is also real, the Fourier coefficients obey $\widehat{\rho}(-\vec{k}) = \widehat{\rho}^*(\vec{k})$.

The convolution theorem now gives the Fourier transform of the smoothed density:

$$\widehat{\rho}_s(\vec{k}) = \widehat{\rho}(\vec{k}) \widehat{W}(\vec{k}) \; , \tag{A.4}$$

where $\widehat{W}(\vec{k})$ is the Fourier transform of the smoothing kernel, with $W(\vec{x})$ normalized to unit volume integral so that $\widehat{W}(0) = 1$. For a point particle, $\widehat{W} = 1$. Note that $\widehat{W}(\vec{k})$ is real and has even parity because $W(\vec{x})$ is real and even.

Given the Fourier transform of the smoothed density, the Fourier transform of the smoothed potential follows immediately from the Poisson equation $\nabla^2 \phi = 4\pi \rho_s$:

$$\widehat{\phi}(\vec{k}) = -4\pi \left(\frac{M}{2\pi}\right)^2 \frac{\widehat{\rho}_s(\vec{k})}{k^2} \ . \tag{A.5}$$

When used with $\widehat{\rho}(\vec{k}) = m_i \exp(-i2\pi \vec{k} \cdot \vec{x}_i / M)$ for a point particle at position $\vec{x}_i$, Eq. A.5 gives the Fourier expansion coefficients for the Ewald summation formula[9] for the potential due to a periodic array of point masses. Although we are considering gravity, any other pair potential may be used simply by replacing $k^{-2}$ in Eq. A.5 by the appropriate Green's function.

The final step is to compute the force from the potential: $\vec{F}(\vec{x}) = -\vec{\nabla}\phi(\vec{x})$. The Fourier coefficients of the force vector are

$$\widehat{\vec{F}}(\vec{k}) = -(2\pi/M)i\vec{k}\widehat{\phi}(\vec{k})\widehat{W}(\vec{k}) = 4\pi \left(\frac{M}{2\pi}\right) \frac{i\vec{k}}{k^2} \widehat{\rho}(\vec{k})\widehat{W}^2(\vec{k}) \ . \tag{A.6}$$

An additional convolution by $W(\vec{x})$ has been included because the force must be averaged over the particle, which has a profile given by $W$. The net force may then be computed by summing the Fourier series as in Eq. A.3.

**Appendix A.2.** *Approximate Forces from a Grid*

With enough terms in the Fourier series, we could evaluate the force with arbitrary precision. However, this would be more costly than a direct summation of the forces in the spatial domain, unless we truncate the sum over wavenumbers and then use a fast transform technique. The FFT algorithm will give an approximate solution to the Poisson equation at $M^3$ grid points in $O(M \log M)^3$ operations, compared with $O(N^2)$ operations for a direct summation in the spatial domain for $N$ particles. The speed of the FFTs is the primary motivation for the PM algorithm, but a penalty is paid in force accuracy, as we show in this subsection.

To analyze the force errors we analyze each step of the force calculation. First, the density assignment of Eq. A.2 remains exact with a grid (provided that $W$ is the appropriate interpolation function) although now the smoothed density is evaluated only at a set of discrete grid points $\vec{x} = \vec{n} = (n_1, n_2, n_3)$, where $(n_1, n_2, n_3) \in [0, M)$ are integers. The smoothing by $W$ is absolutely necessary for a finite number of particles and grid points and it is accomplished in practice using an interpolation scheme such as NGP or TSC.

Next, the volume integral in Eq. A.3 is replaced by a sum over the grid points $\vec{n}$ resulting in a discrete Fourier Transform which may be evaluated using the FFT algorithm. The FFT of the density is equivalent to a sum of the true (continuous) Fourier Transform over Brillouin zones:

$$\widehat{\rho}_{gs}(\vec{k}) = \sum_{\vec{b}} \widehat{\rho}_s(\vec{k} + M\vec{b}) \ , \tag{A.7}$$

where the subscript $g$ indicates that a spatial grid has been used and $\vec{b} = (b_1, b_2, b_3)$ is a triplet of all integers, positive, negative, and zero. Each value of $\vec{b}$ corresponds to one of the Brillouin zones of a periodic lattice. It is important to note that the exact Fourier transform, $\widehat{\rho}_s(\vec{k})$, is defined on an infinite grid of wavevectors. The sum over Brillouin zones in Eq. A.7 therefore represents an aliasing error: high-frequency Fourier components are folded into the first Brillouin zone. If $\widehat{\rho}_s(\vec{k})$ declines rapidly with increasing $|\vec{k}|$, the aliasing error may be small. The smoothing by $W(\vec{x})$ reduces the error if the width of the smoothing kernel is more than a grid spacing, at the expense of a loss of spatial resolution. Our goal is to minimize the aliasing errors in the force for a fixed spatial resolution. This requires that we analyze the rest of the PM algorithm.

The second phase of the PM force calculation is to compute the force on a grid from the FFT of the density. The results are given essentially by Eqs. 2 and 4, though here we allow for a general gradient operator $\vec{D}$:

$$\widehat{\vec{F}}_g(\vec{k}) = 4\pi \left(\frac{M}{2\pi}\right) \vec{D}(\vec{k}) G(\vec{k}) \widehat{\rho}_{gs}(\vec{k}) . \tag{A.8}$$

For exact forces, with $\widehat{\rho}_{gs}$ replaced by $\widehat{\rho}$, we have $G(\vec{k}) = k^{-2}$ and $\vec{D}(\vec{k}) = i\vec{k}$. When a grid is used to evaluate the density, a different choice for $G$ and $\vec{D}$ may be preferable. In general $G$ should be real and even $[G(-\vec{k}) = G(\vec{k})]$ while $\vec{D}$ should be imaginary and odd $[\vec{D}(-\vec{k}) = -\vec{D}(\vec{k})]$ in order that $F_g(\vec{x})$ be real. After evaluating the force in the Fourier domain, it is transformed back to the spatial domain using the FFT:

$$\vec{F}_g(\vec{n}) = M^{-3} \sum_{\vec{k}}{}' \exp(i2\pi\vec{k}\cdot\vec{n}/M) \widehat{\vec{F}}_g(\vec{k}) . \tag{A.9}$$

The primed sum is taken over only the wavevectors in the fundamental Brillouin zone, with components bounded by the Nyquist frequency $\pi/M$: $-M/2 \le k_1, k_2, k_3 < M/2$.

Using Eqs. A.2, A.7, and A.8, Eq. A.9 may be written as a sum over all wavevectors (in all Brillouin zones), demonstrating the errors introduced by a grid:

$$\vec{F}_g(\vec{n}) = M^{-3} \sum_{\vec{k}} \exp(i2\pi\vec{k}\cdot\vec{n}/M) 4\pi \left(\frac{M}{2\pi}\right) \vec{D}(\vec{k}_0) G(\vec{k}_0) \widehat{\rho}(\vec{k}) \widehat{W}(\vec{k}) . \tag{A.10}$$

This is identical to the exact force evaluated at $\vec{x} = \vec{n}$, except that the Green's function and gradient operators are evaluated not at the correct wavevector $\vec{k}$, but at the reduced wavevector $\vec{k}_0$ lying in the fundamental Brillouin zone, with components $k_{0q} = \mathrm{mod}[k_q, M]$. The high frequency components of the force, with wavevectors lying outside of the fundamental Brillouin zone, are incorrect.

On top of these errors, we must still interpolate the force from the grid back to the particles. This interpolation is generally performed with a convolution sum similar to the initial interpolation of the mass density:

$$\vec{F}_{gg}(\vec{x}) = \sum_{\vec{n}} W(\vec{x} - \vec{n}) \vec{F}_g(\vec{n}) . \tag{A.11}$$

A second subscript $g$ has been added to indicate the use of the grid for a second time. Eq. A.11 may also be written in Fourier transform space, with the result

$$\widehat{\vec{F}}_{gg}(\vec{k}) = 4\pi \left(\frac{M}{2\pi}\right) \vec{D}(\vec{k}_0) G(\vec{k}_0) \widehat{\rho}_{gs}(\vec{k}_0) \widehat{W}(\vec{k}) . \tag{A.12}$$

This is equivalent to Eq. (8-19) of Ref. 2.

**Appendix A.3.** *Optimal Anti-Aliasing Filter*

The net effect of introducing a grid into the force calculation is apparent in the comparison of Eqs. A.6 and A.12. The density and one of the smoothing windows are aliased, and the Green's function and gradient operators are evaluated only in the fundamental Brillouin zone, with wavenumber components bounded by the Nyquist frequency. These differences cause force errors.

We would like to make the errors as small as possible by a judicious choice of the Green's function, gradient operator, and smoothing window. This optimization is performed by minimizing the mean squared force error produced at $\vec{x}$ due to a source particle at $\vec{x}_1$, averaging over both the source position $\vec{x}_1$ and the test position $\vec{x}$:

$$\int d^3x_1 \int d^3x \, |\vec{F}_{gg}(\vec{x}) - \vec{F}(\vec{x})|^2 = M^{-3} \sum_{\vec{k}} \int d^3x_1 \, |\widehat{\vec{F}}_{gg}(\vec{k}) - \widehat{\vec{F}}(\vec{k})|^2 \ . \quad (A.13)$$

One of the volume integrals has been converted to a Fourier series using Parseval's theorem. The dependence on $\vec{x}_1$ arises through $\widehat{\rho}(\vec{k}) \propto \exp(-i2\pi\vec{k} \cdot \vec{x}_1/M)$. The sum over wavevectors may be split into a sum over Brillouin zones $\vec{b}$ and a sum over wavevectors $\vec{k}_0$ in the fundamental zone.

We perform the optimization by varying Eq. A.13 with respect to $G(\vec{k}_0)$ for each wavevector in the fundamental Brillouin zone, holding fixed the gradient operator $\vec{D}(\vec{k}_0)$ and the interpolation window $\widehat{W}(\vec{k})$. However, we know that we cannot achieve a good match to the pure inverse square law for a point mass. The finite mesh prevents us from resolving the pair potential for separations smaller than a grid spacing. Worse still, the force between close pairs depends on the orientation of the pairs relative to the mesh, effectively adding small-scale noise to the force law. As Eq. A.12 shows, the noise arises because of aliasing into and out of the fundamental Brillouin zone.

The small-separation force scatter can cause serious problems such as artificially heating a system and producing energy conservation errors. To reduce the scatter we must sacrifice some resolution. We do this by least squares minimization of Eq. A.13, choosing the true pair force $\vec{F}$ to arise from a cloud with shape given by some reference smoothing window that we denote (in the spectral domain) $\widehat{W}_r(\vec{k})$. For example, we may wish to approximate the force from a cloud with a Gaussian profile $W_r(\vec{x})$. This desired shape is to be distinguished from the interpolation window $W(\vec{x})$, which describes our method for discretizing the density and force on a grid and is not a Gaussian. (See, e.g., Eq. 5 for $W$ in the case of TSC interpolation.)

The least-squares optimal Green's function for the force follows from writing the sum over wavevectors in Eq. A.13 as a sum over Brillouin zones and over the wavevectors in the fundamental zone and then differentiating with respect to the Green's function in the fundamental zone. The result is (cf. Eq. [8-22] of Ref. 2)

$$G(\vec{k}_0) = \frac{-i\vec{D}(\vec{k}_0) \cdot \vec{A}(\vec{k}_0)}{|\vec{D}(\vec{k}_0)|^2 \, B^2(\vec{k}_0)} \ , \quad (A.14)$$

where we have defined

$$\vec{A}(\vec{k}_0) \equiv \sum_{\vec{b}} \frac{(\vec{k}_0 + \vec{b}M)}{|\vec{k}_0 + \vec{b}M|^2} \widehat{W}^2(\vec{k}_0 + \vec{b}M)\widehat{W}_r^2(\vec{k}_0 + \vec{b}M) , \quad B(\vec{k}_0) \equiv \sum_{\vec{b}} \widehat{W}^2(\vec{k}_0 + \vec{b}M) .$$
(A.15)

Comparing Eqs. 2 and 4 with Eq. A.8, we see that we have found the optimal anti-aliasing filter $T(\vec{k}) = k^2 G(\vec{k})$.

### Appendix A.4. *Result for TSC Interpolation*

The optimal anti-aliasing filter depends on the interpolation window $W(\vec{x})$, the gradient operator $\vec{D}(\vec{k})$, and on the reference particle shape $W_r(\vec{x})$. To reduce the force scatter to below about 2% rms we use the TSC interpolation window, the gradient operator $\vec{D} = i\vec{k}$, and a linear reference window (particle shape) with $W_r(r) = (24/\pi a^4)(a - 2r)$ for $2r < a$ and $W_r = 0$ otherwise [shape function $S_2(r)$ of Ref. 2]. Hockney and Eastwood (Ref. 2) give the Fourier transforms of the three-dimensional TSC and $S_2$ window functions,

$$\widehat{W}(\vec{k}) = \prod_{q=1}^{3} \left( \frac{2}{k_q} \sin \frac{k_q}{2} \right)^3 , \quad \widehat{W}_r(k) = \frac{12}{(ka/2)^4} \left( 2 - 2\cos \frac{ka}{2} - \frac{ka}{2} \sin \frac{ka}{2} \right) ,$$
(A.16)

where $k = |\vec{k}|$ and the smoothing radius $a$ is expressed in units of the grid spacing. With $a = 3.3$ the scatter in the pair force is at most about 2% for separations about 1 grid spacing; this scatter is reduced to 1% with $a = 3.7$.

Computing the auxiliary quantities $\vec{A}$ and $B$ used in the anti-aliasing filter requires summing over Brillouin zones. In principle these sums should be taken over all $\vec{b}$ to achieve the best results. Fortunately, only a few aliases need be taken (we sum over 5 aliases per dimension) because $\widehat{W}$ and $\widehat{W}_r$ decline fairly rapidly with $k$. However, using shape function $S_2(r)$, the sum for $B$ may be done in closed form, yielding[2]

$$B(\vec{k}_0) = \prod_{q=1}^{3} \left( 1 - \sin^2 \frac{k_q M}{2} + \frac{2}{15} \sin^4 \frac{k_q M}{2} \right) ,$$
(A.17)

where $M$ is the size of the grid.

The optimal filter is computed once at the beginning of a PM simulation and then saved. The computation is easy to parallelize as each virtual processor may be assigned to a grid point in Fourier space and no communication is required between data elements.

### Appendix B   Performance Analysis

In this appendix we provide some details about the performance of our PM code on the CM-5. We also discuss some of our optimization techniques.

### Appendix B.1. *Operation Counts*

The operations counts for each of the phases of the PM are

|  | Flops | Integer Ops | Scan | Communication |
|---|---|---|---|---|
| Density Assignment | $303N$ | $972 FracOC*N$ | $28N$ | $57 FracOC*N$ |
| Force Interpolation | $990N$ | $2916 FracOC*N$ | $84N$ | $171 FracOC*N$ |
| FFT | $10M^3 \log_2 M^3$ | | | |

where $N$ is the number of particles, $M^3$ is the number of grid points, and $FracOC*N$ is the total number of mesh cells which have at least one particle in them. The integer operations are for index calculation. The FFT is called four times, one forward transform and three inverse transforms, for each time step. The force interpolation routine is called three times for each time step, once for each component of the force. The operation counts include these repeat calls. The dependence on $FracOC$ reflects the fact that highly clustered particle distributions have shorter run times.

The effective rates for each of these operations on the CM-5 are

|  | Rate per processing node (PN) |
|---|---|
| Floating Point Ops | 20 MFlops/sec/PN |
| Integer Ops | 20 MOps/sec/PN |
| Scan | 1.0 MOps/sec/PN |
| Communication | 0.125–1.25 MWords/sec/PN |
| FFT | 12.0 MFlops/sec/PN |

We emphasize that these are *effective* rates that are observed in the PM application. For instance, the FFT rate can be decomposed into a computation rate and a communication rate. Since the FFT is an atomic operation for this application, it is most convenient to report a single effective flop rate. Similarly, the communication rate is a composite of a rate for moving data around on a single node, and a rate for moving data between nodes. In addition, there is some overhead associated with each communication operation, associated with determining which pieces of data have to be sent to which processors. The communication rates we quote are a composite of these factors, and are sensitive to the amount of non-locality in the data set, the higher rate corresponding to 100% locality.

### Appendix B.2. *Optimizations*

We performed three distinct optimizations which significantly improved the performance of PM. While the spirit of these optimizations is not specific to the CM-5, the actual implementation may be.

Our first optimization was to reduce the amount of time spent in index calculation. In the inner loop of both the density assignment and the force interpolation phase, we send data from the particle list to the mesh or get data from the mesh to the particle list. In both cases, we have three indices $(I1, I2, I3)$ which must be combined into a single machine index referencing the mesh. This index calculation requires, among other things, determining which indices correspond to which physical processor. Because the CM-5 run-time system allows for a quite general mapping of arrays to processors, this index calculation is quite expensive. However, in our code we are using a simple mapping of the array to the processors. Therefore, it is faster to translate the three indices explicitly into a single Index, and then to do the get or send using that index. In order for this to work, we must be able to reference the mesh as a large 1-D array of length $M^3$ rather than as an $M \times M \times M$ array. On shared memory computers, this is easily accomplished with the FOR-TRAN EQUIVALENCE statement. On the CM-5, this capability is provided with the ALIAS feature described in the CMFortran Utility Library Manual.

The second optimization step was motivated by the fact that the amount of index calculation and the amount communication is proportional to $FracOC*N$, which can be much less than $N$. Consider the force interpolation phase. The FORTRAN 90 code to get the force from the mesh to the head of each segment is

```
WHERE(Segment) FPart = FMesh(Index)
```

where FPart and Index are arrays of size $N$, and FMesh is the force on the mesh. (Recall from the paragraph above that we are indexing this mesh as a one-dimensional vector for this step.) Although only $FracOC*N$ words move from FMesh to FPart, the overhead for this statement is $O(N)$. That is because, for each element of Index a lot of computation is done, and only after that computation completes does the code examine Segment to determine whether that element should execute a get or not. Since $FracOC*N$ may be much less than $N$, the overhead can turn out to be a significant part of the cost of the index calculation and the communication.

We can reduce the cost of the overhead to $O(FracOC*N)$ by packing the indices into an array of size $FracOC*N$ in an intermediate step. We first construct an index array, I0, of length $FracOC*N$ which has a pointer to the head of each segment in FPart. The cost of constructing this is $O(N)$, but is amortized over the 27 iterations we need for our TSC interpolation scheme. In addition, we make Index of size $FracOC*N$ (we need only one index per occupied mesh cell). Then, the get from the mesh to the particles is replaced by

```
FTemp = FMesh(Index)
FPart = FTemp(I0)
```

where FTemp is of length $FracOC*N$. In the first line, $FracOC*N$ elements get from the mesh. Since Index is of length $FracOC*N$, the overhead for this is only $O(FracOC*N)$. In the next step, $FracOC*N$ elements send to the full particle set. Once again, this step is only $O(FracOC*N)$. Consequenlty, by we have reduced the cost of the communication to $O(FracOC*N)$. The fact that the amount of index computation is reduced to $O(FracOC*N)$ and the fact that we perform two communication operations, rather than one single one, are both reflected in the operation counts above.

Although we described this in terms of the force interpolation phase, the same applies to the density assignment phase.

The final optimization we use is to sort the particles so that as much as possible of the data motion is local to a processor. Although the particles and the mesh are in two different data structures, we would like particles to be stored on the same physical processor which stores their NGP. Exact coincidence is not possible because the particles are not necessarily homogeneously distributed: some mesh cells have more particles than others.

In Section 6, we noted that the index we use for sorting the particles is not unique. We exploit this fact by constructing an index which is ordered in the same way that the mesh cells are ordered on the processors. This ordering is specific to the CM-5, of course. But the places in the code where this ordering occurs are isolated, and can easily be modified for other machines, or for an alternative ordering scheme on the CM-5.